\documentclass[12pt]{article}
  \usepackage{amsfonts}
  \usepackage{amsmath}
\usepackage{amssymb}
\usepackage{amscd}
\usepackage[dvips]{graphicx}
\begin{document}
~~
\bigskip
\bigskip
\begin{center}
{\Large {\bf{{{Deformation of nonrelativistic space-time and forces
noticed  by noninertial observer  }}}}}
\end{center}
\bigskip
\bigskip
\bigskip
\begin{center}
{{\large ${\rm {Marcin\;Daszkiewicz}}$ }}
\end{center}
\bigskip
\begin{center}
{ {{{Institute of Theoretical Physics\\ University of Wroc{\l}aw pl.
Maxa Borna 9, 50-206 Wroc{\l}aw, Poland\\ e-mail:
marcin@ift.uni.wroc.pl}}}}
\end{center}
\bigskip
\bigskip
\bigskip
\bigskip
\bigskip
\bigskip
\bigskip
\bigskip
\begin{abstract}
We consider the nonrelativistic particle moving on noncommutative
space-time in the presence of constant force $\vec{F}$. Further,
following the paper M. Daszkiewicz, C.J. Walczyk, Phys. Rev. D  77,
105008 (2008); arXiv: 0802.3575 [mat.ph], we recall that the
considered noncommutativity generates additional force terms, which
appear in the corresponding Newton equation. We demonstrate that the
same force terms can be generated by the proper noninertial
transformation of classical nonrelativistic space-time.
\end{abstract}
\bigskip
\bigskip
\bigskip
\bigskip
\bigskip
\bigskip
\bigskip
\bigskip
\bigskip
 \eject

The suggestion to use noncommutative coordinates goes back to
Heisenberg and was firstly  formalized by Snyder in \cite{snyder}.
Recently, there were also found formal  arguments based mainly  on
Quantum Gravity \cite{2}, \cite{2a} and String Theory models
\cite{recent}, \cite{string1}, indicating that space-time at Planck
scale  should be noncommutative, i.e. it should  have a quantum
nature. Consequently, there appeared a lot of papers dealing with
noncommutative classical and quantum  mechanics (see e.g.
\cite{mech}, \cite{qm}) as well as with field theoretical models
(see e.g. \cite{prefield}, \cite{field}), in which  the quantum
space-time is employed.

In accordance with the Hopf-algebraic classification of all
deformations of relativistic \cite{clas1} and nonrelativistic
\cite{clas2} symmetries, one can distinguish three basic types
of space-time noncommutativity:\\
\\
{\bf 1)} The canonical (soft) deformation
\begin{equation}
[\;{ x}_{\mu},{ x}_{\nu}\;] = i\theta_{\mu\nu}\;, \label{noncomm}
\end{equation}
with constant and antisymmetric tensor $\theta_{\mu\nu}$. The
explicit form of corresponding Poincare Hopf algebra has been
provided in \cite{oeckl}, \cite{chi}, while  its nonrelativistic
limit  has been
proposed in \cite{daszkiewicz}. \\
\\
{\bf 2)} The Lie-algebraic  case 
\begin{equation}
[\;{ x}_{\mu},{ x}_{\nu}\;] = i\theta_{\mu\nu}^{\rho}x_{\rho}\;,
\label{noncomm1}
\end{equation}
with  particularly chosen constant coefficients
$\theta_{\mu\nu}^{\rho}$. Particular kind of such  space-time
modification has been obtained as representations of
$\kappa$-Poincare \cite{kappaP1}, \cite{kappaP2} and
$\kappa$-Galilei \cite{kappaG} Hopf algebras. Besides, the
Lie-algebraic twist deformations of relativistic and nonrelativistic
symmetries have been provided in \cite{lie1}, \cite{lie2} and
\cite{daszkiewicz}, respectively.
\\
\\
{\bf 3)} The quadratic deformation
\begin{equation}
[\;{ x}_{\mu},{ x}_{\nu}\;] =
i\theta_{\mu\nu}^{\rho\tau}x_{\rho}x_{\tau}\;, \label{noncomm2}
\end{equation}
with constant coefficients $\theta_{\mu\nu}^{\rho\tau}$. Its
Hopf-algebraic realization was proposed in \cite{qdef}, \cite{paolo}
and \cite{lie2}  in the case of relativistic symmetry, and in
\cite{daszkiewicz2} for its nonrelativistic counterpart.

Recently, in paper \cite{daszmech}, there has been investigated the
impact of  mentioned above space-time deformations (with commuting
time direction) on a dynamics of simplest classical system - the
nonrelativistic particle moving in a field of constant force
$\vec{F}$\footnote{We consider in \cite{daszmech} four
noncommutative space-times (see (\ref{canps}), (\ref{ssslie1}),
(\ref{hlie2a}) and (\ref{qajj})), which represent the mentioned
above classes of quantum spaces. All of them correspond to the
proper quantum Galilei groups provided in \cite{daszkiewicz} and
\cite{daszkiewicz2} respectively.}. Particulary, it has been
demonstrated that for hamiltonian function
\begin{eqnarray}
H(\vec{p},\vec{x}) = \frac{\vec{p}^2}{2m} +
V(\vec{x})\;\;\;;\;\;\;V(\vec{x}) =
\sum_{i=1}^{3}F_ix_i\;\;\;,\;\;\; F_i = {\rm const.}\;,\label{ham}
\end{eqnarray}
in the case of canonically deformed phase space\footnote{ The
relation between commutator and classical Poisson bracket is given
by $\{\,{ a},{ b}\,\}\to \frac{1}{i\hbar} \,[\,{\hat a},{\hat b}\,]$
$(\hbar =1)$. It should be also noted that all provided below phase
spaces satisfy Jacobi identity.}
\begin{eqnarray}
\{\,x_i,x_j\,\} =\theta_{ij}\;\;\;,\;\;\;
\{\,p_i,p_j\,\}=0\;\;\;,\;\;\;\{\,x_i,p_j\,\}=\delta_{ij}\;,
\label{canps}
\end{eqnarray}
the corresponding Newton equation remains undeformed\footnote{We
find all equations of motion (for all deformed phase spaces)
following the standard procedure \cite{sym}.},\footnote{Due to the
linearity of equations (\ref{eq1}), (\ref{lie1newton1}),
(\ref{slie2newton0}) and (\ref{xxxlie2newton1}) with respect
(quantum) spatial $(x_i)$ directions, their  form is the same on
noncommutative as well as on commutative space-time. Hence, we pass
with the above equations of motion to the classical space without
any their modification.}
\begin{equation}
m\ddot{x}_i  =  F_i\;. \label{eq1}
\end{equation}
In other words, it has been indicated  that in such a case the
space-time noncommutativity (\ref{noncomm}) does not generate any
additional force term.

More interesting situation appears for the Lie-algebraic
modification of nonrelativistic space-time (\ref{noncomm1}). Then,
in the case of the following phase space ($\rho$, $\tau$ -
fixed)\footnote{$x_0 = ct$.}
\begin{eqnarray}
\{\,x_i,x_j\,\} = \frac{1}{\kappa} t(\delta_{i\rho}\delta_{j\tau} -
\delta_{i\tau}\delta_{j\rho})\;\;\;,\;\;\;
\{\,p_i,p_j\,\}=0\;\;\;,\;\;\;\{\,x_i,p_j\,\}=\delta_{ij}\;,
\label{ssslie1}
\end{eqnarray}
with two spatial directions commuting to time direction $t$, we
get\\
\begin{equation}
\left\{\begin{array}{rcl} m\ddot{x}_i  &=&  F_i\;
\\&~~&~\cr
m\ddot{x}_\rho &=& -\frac{m}{\kappa} F_\tau +F_\rho\;\\&~~&~\cr
m\ddot{x}_\tau &=&\frac{m}{\kappa} F_\rho
+F_\tau\;,\end{array}\right.\label{lie1newton1}
\end{equation}
\\
with index $i$ different then $\rho$ and $\tau$. Hence, we see that
in such a case  there appear additional constant force terms
associated with deformation parameter $\kappa$. Similarly,  for the
second Lie-algebraically deformed phase space ($k$, $l$, $\gamma$ -
fixed and different, $i,j \ne \gamma$)
\begin{eqnarray}
&&\{\,x_k,x_\gamma\,\} = \frac{1}{\hat \kappa} x_l\;\;\;,\;\;\;
 \{\,x_l,x_\gamma\,\} = -\frac{1}{\hat \kappa} x_k\;\;\;,\;\;\;\{\,x_k,x_l\,\} = 0\;\;\;,\;\;\;
\{\,p_k,x_\gamma\,\} = \frac{1}{\hat \kappa} p_l\;,\label{hlie2a} \\
&&\{\,p_l,x_\gamma\,\} = -\frac{1}{\hat \kappa}
p_k\;,\;\{\,x_i,p_j\,\}
=\delta_{ij}\;,\;\{\,x_\gamma,p_\gamma\,\}=1\;,\;\{\,p_a,p_b\,\}=0\;,\;a,b=1,2,3
 \;, \nonumber
\end{eqnarray}
with two spatial directions commuting to space, we get the following
modification of Newton equation (\ref{eq1})\\
\begin{equation}
\left\{\begin{array}{rcl} m\ddot{x}_\gamma  &=&  F_\gamma
+\frac{m}{\hat \kappa}F_k\dot{x}_l - \frac{m}{\hat
\kappa}F_l\dot{x}_k\;\\
&~~&~\cr m\ddot{x}_l  &=& F_l+\frac{2m}{\hat
\kappa}F_\gamma\dot{x}_k +{m}\left(\frac{{{F_\gamma}}}{{\hat
\kappa}}\right)^2x_l\;\\
&~~&~\cr m\ddot{x}_k  &=&  F_k -\frac{2m}{\hat
\kappa}F_\gamma\dot{x}_l +{m}\left(\frac{{{F_\gamma}}}{{\hat
\kappa}}\right)^2x_k
 \;.
\label{slie2newton0}\end{array}\right.
\end{equation}
\\
The above result means, that the space-time noncommutativity
(\ref{hlie2a}) generates additional position and velocity dependent
force terms (\ref{slie2newton0}).

Finally, it should be noted, that the (last) quadratically deformed
phase space considered in \cite{daszmech} (with $k$, $l$, $\gamma$
fixed and different, $i,j \ne \gamma$ and $a,b=1,2,3$)
\begin{eqnarray}
&&\{\,x_k,x_\gamma\,\} = \frac{1}{\bar \kappa}tx_l\;\;\;,\;\;\;
\{\,x_l,x_\gamma\,\} = -\frac{1}{\bar
\kappa}tx_k\;\;\;,\;\;\;\{\,x_k,x_l\,\} = 0\;\;\;,\;\;\;
\{\,p_k,x_\gamma\,\} = \frac{1}{\bar \kappa}tp_l\;,\nonumber \\
&&\{\,p_l,x_\gamma\,\} = -\frac{1}{\bar \kappa}
tp_k\;,\;\{\,x_i,p_j\,\}
=\delta_{ij}\;,\;\{\,x_\gamma,p_\gamma\,\}=1\;,\;\{\,p_a,p_b\,\} =0
 \;,\label{qajj}
\end{eqnarray}
 leads to the following
equation of motion\\
\begin{equation}
\left\{\begin{array}{rcl} m\ddot{x}_k  &=&  F_k -\frac{m}{\bar
\kappa}F_\gamma\left(t\dot{x}_l + x_l\right) -\frac{m}{\bar
\kappa}F_\gamma t\left(\dot{x}_l - \frac{1}{\bar
\kappa}F_\gamma tx_k\right)\;\\
&~~&~\cr m\ddot{x}_l  &=&  F_l +\frac{m}{\bar \kappa}F_\gamma
\left(t\dot{x}_k+x_k\right)+\frac{m}{\bar \kappa}F_\gamma
t\left(\dot{x}_k + \frac{1}{\bar \kappa}F_\gamma tx_l\right)\;\\
&~~&~\cr m\ddot{x}_\gamma  &=&  F_\gamma +\frac{m}{\bar
\kappa}F_k\left(t\dot{x}_l +x_l\right)- \frac{m}{\bar
\kappa}F_l\left(t\dot{x}_k+ x_k\right)
 \;.
\label{xxxlie2newton1}\end{array}\right.
\end{equation}\\
Hence, we see, that in the last case there are generated position
and velocity dependent forces as well, but this time,  with time
dependent coefficients (\ref{xxxlie2newton1}).

Obviously, for  deformation parameter $\theta$ running to zero, and
all three  parameters $\kappa$, $\hat{\kappa}$ and $\bar{\kappa}$
approaching   infinity, the
above phase spaces and Newton equations become undeformed.\\

Let us now turn to the more conventional mechanism to generate new
force terms  in Newton equation (\ref{eq1}). First of all, we start
with the classical (commutative) phase space
\begin{eqnarray}
\{\,x_i,x_j\,\} =
\{\,p_i,p_j\,\}=0\;\;\;,\;\;\;\{\,x_i,p_j\,\}=\delta_{ij}\;,
\label{comps}
\end{eqnarray}
and the hamiltonian function (\ref{ham}). Obviously, in such a case
we get the undeformed equation of motion (\ref{eq1}), and  identify
such a system  with the inertial (for example the rest) observer
${\cal{O}}(t,x_1,x_2,x_3)$.

Let us consider the following noninertial transformation from the
observer ${\cal{O}}(t,x_1,x_2,x_3)$ to the nonrelativistic  observer
${\cal{O}}'(t',x_1',x_2',x_3')$
\begin{equation}
\left\{\begin{array}{rcl} t'  &=&  t\;
\\&~~&~\cr
x_i' &=& x_i + v_it + y_i\;\\&~~&~\cr x_\tau' &=& x_\tau +
\frac{1}{2\kappa}F_\rho t^2 + v_\tau t +y_\tau\\&~~&~\cr x_\rho' &=&
x_\rho - \frac{1}{2\kappa}F_\tau t^2 + v_\rho t +y_\rho
\;,\end{array}\right.\label{t1}
\end{equation}
where $v_a$ and $y_a$ $(a=i,\rho,\tau)$ denote arbitrary constants.
As one can easily see, the above transformation connects the
inertial observer ${\cal{O}}(t,x_1,x_2,x_3)$ with the uniformly
accelerated (in directions $\rho$ and $\tau$) observer
${\cal{O}}'(t',x_1',x_2',x_3')$. By simple calculation one can also
check that after transformation (\ref{t1}) the Newton equation
(\ref{eq1})
 takes the form\\
\begin{equation}
\left\{\begin{array}{rcl} m\ddot{x}_i'  &=&  F_i\;
\\&~~&~\cr
m\ddot{x}_\rho' &=& -\frac{m}{\kappa} F_\tau +F_\rho\;\\&~~&~\cr
m\ddot{x}_\tau' &=&\frac{m}{\kappa} F_\rho
+F_\tau\;,\end{array}\right.\label{translie1newton1}
\end{equation}\\
i.e. there appear the (additional) inertial force terms which are
the same as in the equation (\ref{lie1newton1}). Hence, we see that
from such a  point of view, one can identify the dynamical effects
of space-time noncommutativity (\ref{ssslie1}) with the ones
generated by  noninertial transformation (\ref{t1}), while the
deformation parameter $\kappa$ describes the degree of
noninertiality.

Similarly, in the case of phase space deformation (\ref{hlie2a}) and
(\ref{qajj}), one can check that the additional force terms which
appear in  Newton equations (\ref{slie2newton0}) and
(\ref{xxxlie2newton1}), are generated by the following noninertial
transformations of commutative space-time \\
\begin{equation}
\left\{\begin{array}{rcl} t  &=&  t'\;
\\&~~&~\cr
x_\gamma &=& x_\gamma ' -
\frac{1}{\hat{\kappa}}F_k\int_{0}^{t}x_l'(\tau)d\tau +
\frac{1}{\hat{\kappa}}F_l\int_{0}^{t}x_k'(\tau)d\tau \;\\&~~&~\cr
x_l &=& x_l' -
\frac{2}{\hat{\kappa}}F_\gamma\int_{0}^{t}x_k'(\tau)d\tau -
\left(\frac{F_\gamma}{\hat{\kappa}}\right)^2\int_{0}^{t}\int_{0}^{\tau_2}x_l'(\tau_1)d\tau_1d\tau_2
\;\\&~~&~\cr x_k &=& x_k'
+\frac{2}{\hat{\kappa}}F_\gamma\int_{0}^{t}x_l'(\tau)d\tau -
\left(\frac{F_\gamma}{\hat{\kappa}}\right)^2\int_{0}^{t}\int_{0}^{\tau_2}x_k'(\tau_1)d\tau_1d\tau_2
\;,\end{array}\right.\label{t2}
\end{equation}\\
and\\
\begin{equation}
\left\{\begin{array}{rcl} t  &=&  t'\;
\\&~~&~\cr
x_k &=& x_k ' + \frac{1}{\bar\kappa}F_\gamma\int_{0}^{t}\left(\tau
x_l'(\tau)\right)d\tau + \frac{1}{\bar\kappa}F_\gamma
t\int_{0}^{t}x_l'(\tau)d\tau\\&~~&~\cr~&&~~~~~~~~~~~~~~~~~~~~~~~~
-\frac{2}{\bar
\kappa}F_\gamma\int_{0}^{t}\int_{0}^{\tau_2}x_l'(\tau_1)d\tau_1d\tau_2
+ {\cal A}_{\bar\kappa,k}(t)\;\\&~~&~\cr x_l &=& x_l' -
\frac{1}{\bar\kappa}F_\gamma\int_{0}^{t}\left(\tau
x_k'(\tau)\right)d\tau -\frac{1}{\bar\kappa}F_\gamma
t\int_{0}^{t}x_k'(\tau)d\tau\\&~~&~\cr~&&~~~~~~~~~~~~~~~~~~~~~~~~
+\frac{2}{\bar
\kappa}F_\gamma\int_{0}^{t}\int_{0}^{\tau_2}x_k'(\tau_1)d\tau_1d\tau_2
+ {\cal A}_{\bar\kappa,l}(t) \;\\&~~&~\cr x_\gamma &=& x_\gamma'
-\frac{1}{\bar\kappa}F_k\int_{0}^{t}\left(\tau
x_l'(\tau)\right)d\tau +
\frac{1}{\bar\kappa}F_l\int_{0}^{t}\left(\tau
x_k'(\tau)\right)d\tau\;,\end{array}\right.\label{t3}
\end{equation}\\
with $\frac{d^2}{dt^2}{\cal A}_{\bar\kappa,k(l)}(t) = -\frac{1}{\bar
\kappa^2}F_\gamma^2 t^2 x_{k(l)}'$, respectively. It means, that
both effects of  deformations can be identified  with the
consequences of noninertial transformations (\ref{t2}), (\ref{t3}),
which connect the
 observer ${\cal{O}}(t,x_1,x_2,x_3)$ with the observer
 ${\cal{O}}'(t',x_1',x_2',x_3')$. From such point of view the
 deformation parameters ${\hat \kappa}$ and ${\bar \kappa}$ describe the degree of
noninertiality as well.

Of course, for all three  parameters $\kappa$, $\hat{\kappa}$ and
$\bar{\kappa}$ approaching  infinity, the above transformations
become  identity.\\

  In this short article we demonstrate, that the additional
 force terms which appear in  Newton equations  (\ref{lie1newton1}),
(\ref{slie2newton0}) and (\ref{xxxlie2newton1}), can be generated
equivalently  in
 two ways - by the presence of space-time noncommutativity (\ref{ssslie1}), (\ref{hlie2a})
 and (\ref{qajj}), or by the noninertial transformation of classical space (\ref{t1}), (\ref{t2})
 and (\ref{t3}). In such a way we did show that these two approaches
  lead to the same additional force term (\ref{lie1newton1}), (\ref{slie2newton0}) and (\ref{xxxlie2newton1}).

  The above
 results have been obtained for nonrelativistic particle moving in
 the presence of constant force $\vec{F}$, but in principle, it can be extended to an arbitrary potential
 function $V(x)$. However, due to
  the nonlinear form of the  noncommutative space-time
  function $V(x)$, such a generalization seems to be  quite complicated from calculational  point of view.

\section*{Acknowledgments}
The author would like to thank J. Lukierski and Z. Haba
for valuable discussions.\\
This paper has been financially supported by Polish Ministry of
Science and Higher Education grant NN202318534.

\end{document}